\documentstyle[12pt]{article}
\begin{document}

\null\vskip-1cm
\rightline{hep-ph/yymmddd}
\rightline{August 1996}

\vspace{1cm}

\begin{center}

{\large\bf TESTING QCD PREDICTIONS FOR} \\
{\large\bf MULTIPLICITY DISTRIBUTIONS AT HERA}
\vspace{1.cm}

{\sc S. Hegyi}\footnote{E-mail: hegyi@rmki.kfki.hu}

\vspace{.3cm}

{\normalsize
	KFKI Research Institute for \\
	Particle and Nuclear Physics of the \\
	Hungarian Academy of Sciences, \\
	H--1525 Budapest 114, P.O. Box 49. \\
	Hungary
}

\end{center}

\vspace{.9cm}

\begin{abstract}
   On the basis of a recently introduced generalization of the
   negative binomial distribution the influence of
   higher-order perturbative QCD effects on multiplicity fluctuations
   are studied for deep inelastic $e^+p$
   scattering at HERA energies. It is found that the multiplicity
   distributions measured by the H1 Collaboration indicate
   violation of infinite divisibility in agreement with
   pQCD calculations. Attention is called to future experimental
   analysis of combinants whose nontrivial
   sign-changing oscillations are
   predicted using the generalized negative binomial law.
\end{abstract}

\newpage
\pagestyle{plain}
\noindent
Over the past 15 years of experimental and theoretical study of multiplicity
distributions (MDs) in a variety of collision processes a universal law
emerged known as the negative binomial (NB) regularity~[1]. The negative
binomial distribution (NBD) proved to be very successful in describing
the observed shape of MDs not only in full phase-space but also in
restricted subdomains. At the same time the NBD naturally arises in branching
processes which share many common features with the parton cascades of QCD.
Thus the theoretical basis of the NB regularity is well founded, at least
qualitatively.

The NBD has two parameters, the shape parameter $k$ and the
mean value $\langle n\rangle$. Its analytic form can be written as
$$
	P_n=\frac{1}{n!}\frac{\Gamma(k+n)}{\Gamma(k)}
	\left(\frac{\theta}{1+\theta}\right)^nP_0\eqno(1)
$$
with
$$
	P_0=(1+\theta)^{-k}\quad\mbox{and}\quad
	\theta=\frac{\langle n\rangle}{k}.\eqno(2)
$$
The probability generating function ${\cal G}(z)=\sum_{n=0}^\infty P_n z^n$
has the form
$$
	{\cal G}(z)=(1+(1-z)\,\theta\,)^{-k}.\eqno(3)
$$
A fundamental property of the distribution is that asymptotically $P_n$
exhibits a nice scaling behaviour~[2]
$$
        \lim_{
        n\rightarrow\infty,\,\langle n\rangle\rightarrow\infty
        \atop n/\langle n\rangle\mbox{\ \tiny fixed}}
        P_n=\frac{1}{\langle n\rangle}\,
        \psi\left(\frac{n}{\langle n\rangle}\right)
        \eqno(4)
$$
which is the famous KNO scaling law~[3]. The preasymptotic MDs
can be reconstructed from the asymptotic scaling function
$\psi(z=n/\bar n)$ via Poisson transform:
$$
        P_n=\int_0^\infty \psi(z)\,\frac{(\bar{n}z)^n}{n!}
        e^{-\bar{n}z}\,dz.\eqno(5)
$$
For the NBD $\psi(z)$ turns out to be a gamma distribution.
In fact Eq.~(5) is a
special superposition of Poisson distributions, known as
Poisson mixing in the mathematical literature~[4].
The superposition is such that the mean values $\langle n\rangle$ of the
Poisson components vary according to
$\psi(z)$ which is called the mixing distribution.
According to the theorem of Ottestad and Consael~[4]
the resulting discrete distribution $P_n$ has probability
generating function equivalent to the characteristic function of $\psi(z)$
and, consequently, the factorial moments(cumulants)  of $P_n$ are equivalent
to the ordinary moments(cumulants) of $\psi(z)$.

During the last few years discrepancies have been found between the NBD and
the MDs measured at the highest available energies. This is not
surprising; it is very unlikely that a two-parameter discrete distribution
can describe all the details of MDs in $ll$, $lh$ and $hh$ collisions.
Besides the experimentally found deviations, the perturbative QCD calculations
also indicate that the NB regularity is not general enough to account
higher-order pQCD effects~[6]. These manifest in the violation of infinite
divisibility of the MDs in sharp contradiction to the NBD being infinitely
divisible. The analysis of experimental data confirmed the pQCD
calculations~[7,8].

In a recent paper we have generalized the NBD by incorporating
some pQCD-based characteristics of MDs~[9]. The model involves
an additional parameter, $\mu$, with the following physical meaning.
Let us preserve the asymptotic scaling
form of the NBD (i.e. the gamma distribution) but not simply
in the scaling variable~$z$, rather,
in a certain power of it, say in~$z^\mu$
with $\mu>0$. Then the asymptotic scaling function can be written as
$$
        \psi(z)=\frac{\mu}{\Gamma(k)}\,\lambda^{\mu k}
        z^{\mu k-1}\exp\left(-[\lambda z]^{\mu}\right) \eqno(6  )
$$
which is the generalized gamma distribution~[5]. The ordinary gamma
distribution is the $\mu=1$ special case. Obviously, the $q$th moment
of the modified $\psi(z)$ involves the fractional rank $q/\mu$.
Due to the theorem of Ottestad and Consael
quoted before, the same rescaled rank $q/\mu$ appears in the
$q$th factorial moment of $P_n$ defined by
the Poisson transform of the modified
$\psi(z)$. Thus one can reproduce
possible enhancement ($\mu<1$) or suppression ($\mu>1$) of multiplicity
fluctuations with respect to the NBD by fitting the Poisson
transform of the generalized gamma distribution
to the observed $P_n$.

The Poisson transform of Eq.~(6) can be expressed in terms of Fox's
generalized hypergeometric function~[9]. Without going into the details
we recapitulate the final result:
$$
   P_n=
   \frac{1}{n!\,\Gamma(k)}\;{\sf H}^{1,1}_{1,1}
   \left[
      \,\frac{1}{\theta}\left|
      \begin{array}{c}
         (1-n,\; 1)             \\
         (k,\; 1/\mu)
      \end{array}
   \right],\right.\;0<\mu<1\eqno(7)
$$
and
$$
   P_n=
   \frac{1}{n!\,\Gamma(k)}\;{\sf H}^{1,1}_{1,1}
   \left[
      \,\theta\left|
      \begin{array}{c}
         (1-k,\; 1/\mu)         \\
         (n,\; 1)
      \end{array}
   \right],\right.\;\mu>1\eqno(8)
$$
where
$$
	\theta=\frac{\langle n\rangle\,\Gamma(k)}
	{\Gamma(k+1/\mu)}\eqno(9)
$$
and ${\sf H}(\cdot)$ denotes the Fox-function.
The probability generating function of $P_n$ is given by
$$
   {\cal G}(z)=
   \frac{1}{\Gamma(k)}\;{\sf H}^{1,1}_{1,1}
   \left[
      \,\frac{1}{t}\left|
      \begin{array}{c}
         (1,\; 1)       \\
         (k,\; 1/\mu)
      \end{array}
   \right],\right.\;0<\mu<1\eqno(10)
$$
and
$$
   {\cal G}(z)=
   \frac{1}{\Gamma(k)}\;{\sf H}^{1,1}_{1,1}
   \left[
      \,t\left|
      \begin{array}{c}
         (1-k,\; 1/\mu)         \\
         (0,\; 1)
      \end{array}
   \right],\right.\;\mu>1\eqno(11)
$$
with $t=(1-z)\,\theta$. For $\mu=1$ the negative binomial distribution
is recovered.
The necessity of two separate expressions for $P_n$ and ${\cal G}(z)$
follows from the existence conditions of ${\sf H}(x)$ discussed in~[9].

The splitted parameter space in~$\mu$ reflects an important difference
between the two expressions for $P_n$ given by Eqs.~(7-8).
According to
the theorem of Bondesson~[10] the generalized gamma density~(6) is
infinitely divisible only for $0<\mu\leq1$. For $\mu>1$ its characteristic
function is entire analytic function of finite order and hence it must have
complex zeroes which is not permitted for infinitely divisible entire
characteristic functions~[10]. Let us recall Maceda's theorem~[4] which
states that a discrete distribution $P_n$ defined by Poisson mixing~(5)
is infinitely divisible if and only if
the mixing distribution $\psi(z)$ with $z\in(0,\infty)$ is infinitely
divisible. Accordingly, in case of Eq.~(7) $P_n$
satisfies the requirements of infinite divisibility,
just as the NBD for $\mu=1$, whereas the $\mu>1$ case given by Eq.~(8)
violates this feature. Thus our simple generalization of the NBD is capable
of reproducing a basic prediction of pQCD calculations.
The additional parameter~$\mu$ measures the
degree of violation of infinite divisibility for $\mu>1$.

To see how Eqs.~(7-8) work in practice we carried out fits to the recent
multiplicity data of the H1 Collaboration measured in deep inelastic
$e^+p$ scattering at HERA over a large kinematic region~[11].
The MDs have been studied in pseudorapidity $\eta^*$ domains of varying
size in the
current fragmentation region of the hadronic centre-of-mass frame.
Comparison of the MDs to the NB and
lognormal (LN) distributions showed that both models give an acceptable
description of the uncorrected data only in the smallest
pseudorapidity domains.
For widening $\eta^*$-intervals the quality of fits becomes
progressively worse~[11].

In our fitting procedure numerical evaluation of the integral defining
the Poisson transform~(5) has been carried out using 96-point gaussian
quadrature. It is turned out that keeping the shape parameter fixed at
$k=1$ and fitting only $\langle n\rangle$ and~$\mu$
produces reasonable description of the H1 data for $P_n$
tabulated in ref.~[11]. Thus our theoretical $P_n$ is the Poisson
transform of the Weibull distribution~[5].
The scale parameter of $\psi(z)$ is restricted to
$\lambda=\Gamma(1/\mu)/\mu$ by the normalization condition
$\int_0^\infty z\,\psi(z)dz=1$~[9]. The results of fits are collected
in Table~1. As is seen the values of the $\chi^2$ are satisfactory, with
the only exception of the fit corresponding to $1<\eta^*<3$ and
$W=80\div115$ GeV. In this case the NBD provides a better
account of the data. Let us observe that $\mu>1$ and increases
for widening $\eta^*$-intervals signalling increasingly dominant
violation of infinite divisibility. In our opinion
this is the reason why the NB and LN
distributions (both being infinitely divisible) produce progressively
worse fits for large pseudorapidity domains. The quality of our fits is
illustrated in Fig.~1 for $1<\eta^*<4$.

In the remaining part of this Letter we consider the question of
infinite divisibility of MDs on the basis of combinants.
According to L\'evy's theorem~[4]
a discrete distribution is infinitely divisible if and only if
its probability generating function can be written in the form
$$
	{\cal G}(z)=\exp(\lambda\,(g(z)-1))\eqno(12)
$$
where $\lambda>0$ and $g(z)$ is another probability generating function.
The discrete distributions having
${\cal G}(z)$ of the form of Eq.~(12) are known also as compound Poisson
distributions. These can be regarded as convolutions of Poisson singlet,
Poisson doublet, Poisson triplet, etc. distributions since Eq.~(12) can be
rewritten as
$$
	{\cal G}(z)=\prod_{q=1}^\infty\exp({\cal C}_q(z^q-1))
	\eqno(13)
$$
i.e. as the product of the generating functions of the Poisson components
having mean values ${\cal C}_q$. The ${\cal C}_q$ are proportional to
the probabilities given by $g(z)$ with constant of proportionality
$\lambda$ hence they can not be negative for infinitely divisible
distributions~[4].

After Kauffmann and Gyulassy~[12]
the quantities ${\cal C}_q$ acquired the name
combinants in multiparticle
phenomenology. They can be expressed in terms of the
probability ratios ${\cal P}_q=P_q/P_0$ according to
$$
        {\cal C}_q={\cal P}_q-{1\over q}
        \sum_{i=1}^{q-1}i\,{\cal C}_i{\cal P}_{q-i}
        \eqno(14)
$$
see~[1,12,13].
Although the requirement $P_0>0$ can be a drawback for full phase-space
analysis of MDs, in restricted subdomains the combinants have a number
of advantages. First of all, as Eq.~(14) shows, the knowledge of
${\cal C}_q$ requires only the finite number of probabilities
$P_{n\leq q}$.
This is extremely useful since testing the violation of infinite
divisibility of $P_n$
can be realized without the possible influence of truncating
the high-multiplicity tail
which can mimic pQCD effects~[1,14]. Furthermore, one need
not know the probabilities~$P_n$ themselves; the combinants can be
determined directly from the unnormalized topological
cross-sections~$\sigma_n$ because they involve only ratios of~$P_n$.
Thus the statistical and systematic uncertainties of
$\sigma_{tot}$
are filtered out by the combinants. They also share some common features
with the factorial cumulants, e.g. ${\cal C}_1=\langle n\rangle$ and
${\cal C}_{q\geq2}=0$ for the Poisson distribution, further,
both quantities should be nonnegative for infinitely divisible
distributions.

Since the generalization of the NBD
obtained by the Poisson transform of Eq.~(6) can
violate infinite divisibility it is of interest to study the
behaviour of its combinants. For $\mu\leq1$ one expects positive
${\cal C}_q$ in all orders but for $\mu>1$ sign-changing oscillations
may appear in the $q$-dependence of ${\cal C}_q$ similarly to the factorial
cumulant-to-moment ratios $H_q$ of the distribution~[9].
We have calculated the combinants of the generalized NBD with the help of
Eq.~(14) using the same numerical integration procedure for the evaluation
of $P_n$ as described earlier. A typical result for the behaviour of
$\log|{\cal C}_q|$ over the
\hbox{$\mu$-$q$}~plane is shown in Fig.~2a for $k=3/2$
and $\langle n\rangle=10$
(with the same $k$ the behaviour of $H_q$ is studied in~[9]).
The peculiar structure for $\mu>1$ is due to
sign-changing oscillations of ${\cal C}_q$. At a fixed~$q$ the neighbouring
bumps are ${\cal C}_q$ intervals of opposite sign in~$\mu$. For $\mu\leq1$
${\cal C}_q$ is always positive as expected. In Fig.~2b slices are shown
with $q=5$, $20$ (left) and with $\mu=1$, $4$ (right). It is seen that
the pattern of oscillations is nontrivial, i.e. not alternating in sign
as~$q$ takes even/odd values. Qualitatively similar sign-changing
oscillations of ${\cal C}_q$ occur for a different choice of
$\langle n\rangle$ and~$k$.

Finally we provide a further example for the usefulness of combinants in the
analysis of MDs. It is often stated that the factorial cumulant-to-moments
raios $H_q$ are very sensitive to tiny details of the high-multiplicity
tail of $P_n$ (and thus to truncation effects as well). Let us consider
this widespread opinion somewhat closer for the NBD. The unnormalized
factorial moments~$\xi_q$ of the NBD are given by
$$
        \xi_q={\Gamma(k+q)\over\Gamma(k)}\,\theta^q
        \eqno(15)
$$
where $\theta$ is given in Eq.~(2). The unnormalized factorial
cumulants~$f_q$ take the form
$$
        f_q=k\,\Gamma(q)\,\theta^q\eqno(16)
$$
hence for $H_q=f_q/\xi_q$ one obtains~[6]
$$
        H_q=k\,{\Gamma(k)\,\Gamma(q)\over\Gamma(k+q)}=
        k\,\hbox{\rm B}(k,q)\eqno(17)
$$
with B$(\cdot)$ denoting the Euler beta-function. The combinants of the NBD
are given by~[15,16]
$$
        {\cal C}_q={k\over q}\left({\theta\over1+\theta}\right)^q.
        \eqno(18)
$$
Comparing Eqs.~(1) and (15) we get
$$
        q!\,{\cal P}_q={\Gamma(k+q)\over\Gamma(k)}
        \left({\theta\over1+\theta}\right)^q={\xi_q\over(1+\theta)^q}
        \eqno(19)
$$
whereas the comparison of Eqs.~(18) and (16) yields
$$
        q!\,{\cal C}_q=k\,\Gamma(q)
        \left({\theta\over1+\theta}\right)^q={f_q\over(1+\theta)^q}.
        \eqno(20)
$$
Thus one arrives at the rather unexpected result
$$
        H_q={{\cal C}_q\over{\cal P}_q}=
	1-{1\over q}\sum_{i=1}^{q-1}
        i\,{\cal C}_i{P_{q-i}\over P_q}
        \eqno(21)
$$
which shows that the factorial
cumulant-to-moment ratios $H_q$ of the NBD are completely
insensitive to the $P_{n>q}$ tail of the distribution
and carry essentially the same information as the combinants.
This is a very special property of
the NBD which does not hold in general for infinitely divisible
distributions.

In summary, we have investigated the violation of
infinite divisibility of MDs. This
particular feature plays a central role in recent
studies of multiplicity
fluctuations since the higher-order pQCD calculations predict
its appearance for MDs measured in
full phase-space and in restricted subdomains.
As a consequence, departures
arise from the  NB regularity.

In a previous paper we have generalized the NBD by
extending the validity of its asymptotic scaling form
to a certain power of the scaling variable.
By this modification
one can reproduce possible suppression/enhancement
of multiplicity fluctuations with respect to the NBD as well as
the violation of infinite divisibility of MDs. Fitting the generalized
NBD to the recent experimental data for $P_n$ measured by the H1
Collaboration in deep inelastic $e^+p$ scattering
at HERA we have obtained good agreement.
According to our results the violation of infinite divisibility of $P_n$
is increasingly dominant in widening pseudorapidity intervals.

Investigating the combinants of the
generalized NBD we have found nontrivial sign-changing oscillations of
these quantities for the subdomain of parameter space violating
infinite divisibility. In our view the higher-order pQCD effects
on multiplicity fluctuations can be
studied most effectively by measuring the combinants in restricted
phase-space volumes. The constraints of conservation laws are less
pronounced, hence the dynamical effects are more visible,
and the sign-changing oscillations of the combinants are
not corrupted by finite statistics effects.
The pQCD prediction for the $q$-dependence of ${\cal C}_q$
(location of the first minimum, etc.) would be very important
to identify properly the dynamics underlying the oscillations.

%It will be interesting to see how can the combinants help our
%understanding of multiparticle production in high-energy collisions
%processes.

\vspace{1.5cm}
\noindent
ACKNOWLEDGEMENTS

\vskip.5cm
\noindent
This work was supported by the Hungarian Science
Foundation under Grant No. OTKA-F4019/1992.

\newpage\null\noindent
{\LARGE\bf References}
\vskip.5cm

\begin{tabbing}
 [00] \=  akarmiakarmiakarmiakarmiakarmiakarmiakarmiakarmiakarmi        \kill
 {[}1]\>  R. Ugoccioni, A. Giovannini and S. Lupia,
	  {\it ``Multiplicity Distributions''\/},			\\
      \>  in Proc. XXIV Int. Symp. on Multiparticle Dynamics, p. 384,   \\
      \>  Eds. A. Giovannini, S. Lupia and R. Ugoccioni,
	  World Scientific, 1995.					\\
 {[}2]\>  P. Carruthers and C.C. Shih,
          {\it Int. J. Mod. Phys. A\/} 2 (1987) 1447.                   \\
 {[}3]\>  Z. Koba, H.B. Nielsen and P. Olesen,
          {\it Nucl. Phys. B\/} 40 (1972) 314.                          \\
 {[}4]\>  N.L. Johnson and S. Kotz, {\it Distributions in Statistics\/},
          {\it Vol. 1.,}                                        	\\
      \>  {\it Univariate Discrete Distributions\/},
          Wiley, 1970.                                  		\\
 {[}5]\>  N.L. Johnson and S. Kotz, {\it Distributions in Statistics\/},
          {\it Vol. 2.,}                                        	\\
      \>  {\it Univariate Continuous Distributions\/},
          Wiley, 1970.                                  		\\
 {[}6]\>  I.M. Dremin, {\it Physics Uspekhi\/} 37 (1994) 715.           \\
 {[}7]\>  I.M. Dremin et al.,
	  {\it Phys. Lett. B\/} 336 (1994) 119.                         \\
 {[}8]\>  SLD Collaboration, K. Abe et al.,
          {\it Phys. Lett. B\/} 371 (1996) 149.                         \\
 {[}9]\>  S. Hegyi, {\it ``Multiplicity Distributions in Strong
	  Interactions:}						\\
      \>  {\it A Generalized Negative Binomial Model''\/},
          hep-ph/9608346, August 1996.      				\\
 {[}10]\> L. Bondesson, {\it Ann. Prob.\/} 7 (1979) 965,
          {\it Scand. Actuar. J.\/} 78 (1978) 48.                       \\
 {[}11]\> S. Aid et al., H1 Collaboration,
	  {\it ``Charged Particle Multiplicities in}			\\
       \> {\it Deep Inelastic Scattering at HERA''\/},
	  DESY Report, DESY-96-160					\\
       \> and hep-ex/9608011, August 1996.				\\
 {[}12]\> S.K. Kauffmann and M. Gyulassy,
	  {\it J. Phys. A\/} 11 (1978) 1715.				\\
 {[}13]\> S. Hegyi, {\it Phys. Lett. B\/} 318 (1993) 642.		\\
 {[}14]\> R. Ugoccioni, A. Giovannini and S. Lupia,
          {\it Phys. Lett. B\/} 342 (1995) 387.				\\
 {[}15]\> S. Hegyi, {\it Phys. Lett. B\/} 309 (1993) 443.		\\
 {[}16]\> I. Szapudy and A.S. Szalay,
	  {\it Astrophys. J.\/} 408 (1993) 43.
\end{tabbing}

\newpage
\pagestyle{empty}

\begin{tabular}{||c|c|c|c|r||}\hline
$\eta^*$-interval & W (GeV) & $\mu$ & $\langle n\rangle$ & $\chi^2/$ d.o.f.\\
\hline\hline
$1<\eta^*<2$ & \ $80\div115$ & $1.96\pm0.08$ &
                               $2.44\pm0.04$ & 18/12 \ \ \ \\ \cline{2-5}
             &  $115\div150$ & $1.82\pm0.07$ &
                               $2.50\pm0.04$ &  5/13 \ \ \ \\ \cline{2-5}
             &  $150\div185$ & $1.87\pm0.08$ &
                               $2.61\pm0.05$ & 12/13 \ \ \ \\ \cline{2-5}
             &  $185\div220$ & $1.84\pm0.09$ &
                               $2.64\pm0.06$ &  7/13 \ \ \ \\ \hline\hline
$1<\eta^*<3$ & \ $80\div115$ & $2.46\pm0.07$ &
                               $4.88\pm0.06$ & 44/16 \ \ \ \\ \cline{2-5}
             &  $115\div150$ & $2.09\pm0.06$ &
                               $5.05\pm0.07$ & 25/17 \ \ \ \\ \cline{2-5}
             &  $150\div185$ & $2.07\pm0.07$ &
                               $5.29\pm0.08$ &  9/19 \ \ \ \\ \cline{2-5}
             &  $185\div220$ & $2.10\pm0.08$ &
                               $5.33\pm0.09$ &  7/20 \ \ \ \\ \hline\hline
$1<\eta^*<4$ & \ $80\div115$ & $3.70\pm0.14$ &
                               $6.42\pm0.06$ & 17/17 \ \ \ \\ \cline{2-5}
             &  $115\div150$ & $3.18\pm0.11$ &
                               $7.02\pm0.07$ & 18/19 \ \ \ \\ \cline{2-5}
             &  $150\div185$ & $2.92\pm0.09$ &
                               $7.49\pm0.08$ & 16/21 \ \ \ \\ \cline{2-5}
             &  $185\div220$ & $2.75\pm0.10$ &
                               $7.67\pm0.09$ & 12/21 \ \ \ \\ \hline\hline
$1<\eta^*<5$ & \ $80\div115$ & $4.66\pm0.23$ &
                               $6.87\pm0.06$ & 12/17 \ \ \ \\ \cline{2-5}
             &  $115\div150$ & $4.44\pm0.18$ &
                               $7.70\pm0.06$ & 26/20 \ \ \ \\ \cline{2-5}
             &  $150\div185$ & $4.16\pm0.16$ &
                               $8.39\pm0.08$ & 19/21 \ \ \ \\ \cline{2-5}
             &  $185\div220$ & $4.08\pm0.18$ &
                               $8.79\pm0.08$ & 16/22 \ \ \ \\ \hline
\end{tabular}

\vskip.5cm\noindent
Table 1. \ Results of fits to the H1 data for $P_n$
with the Poisson transform of the
generalized gamma distribution~(6) for fixed shape parameter
$k=1$ (Weibull case). The errors are only statistical.

\vspace{1.2cm}
\noindent
FIGURE CAPTIONS

\vskip.5cm
\noindent
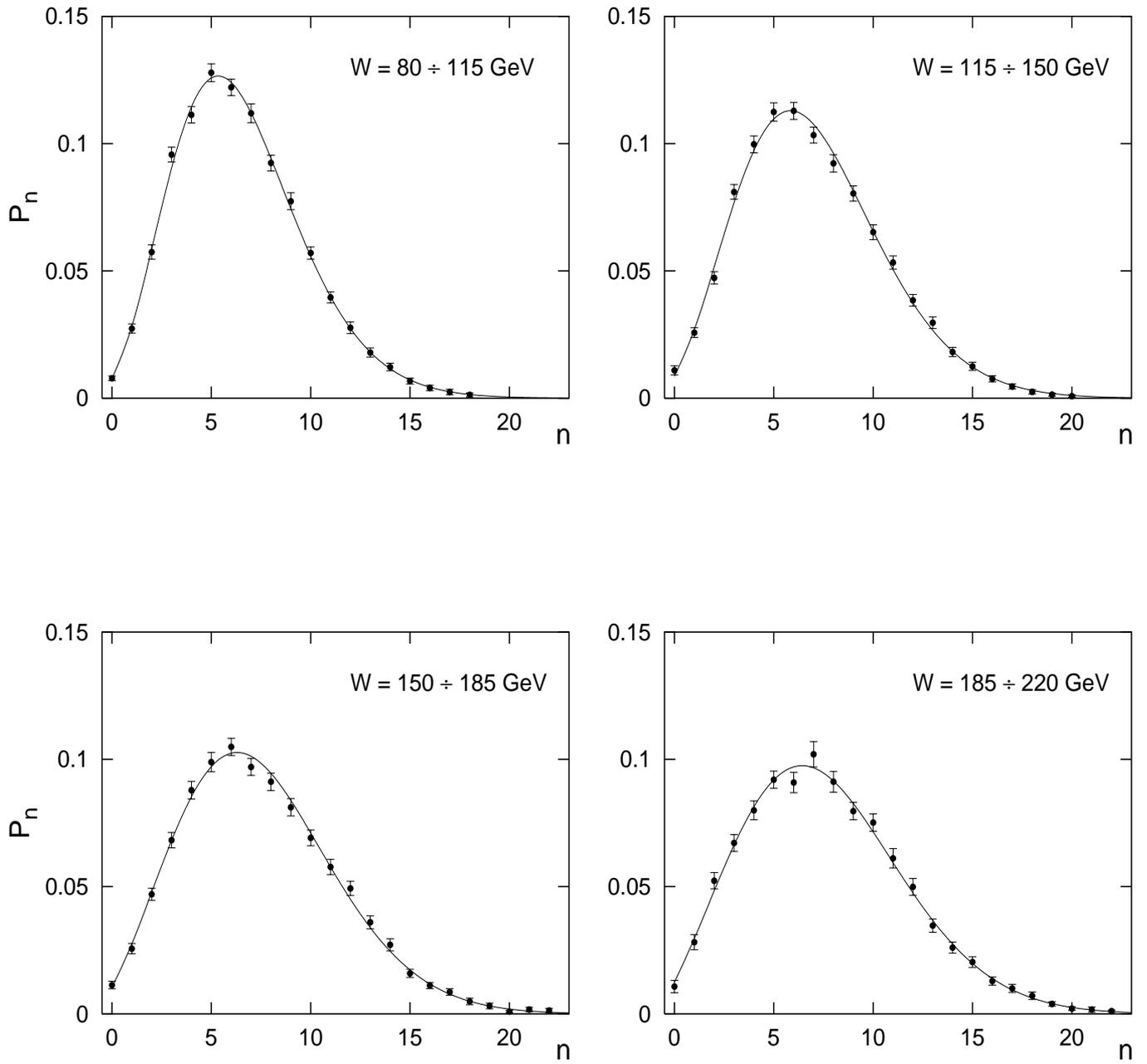
Fig. 1: \ The best-fit theoretical distributions for $1<\eta^*<4$,
see the text for details. The corresponding parameters are collected
in Table~1. The displayed errors are only statistical.

\vskip.5cm
\noindent
Fig. 2a:\ Sign-changing oscillations of the combinants ${\cal C}_q$
of the Poisson transformed generalized gamma distribution.
The parameters kept fixed
are $k=3/2$ and $\langle n\rangle=10$. For clarity only the odd-rank
combinants are displayed.

\vskip.5cm
\noindent
Fig. 2b:\ Slices through Fig.~2a with $q=5$, 20 (left) and with
$\mu=1$, 4 (right).
The $\mu=1$ case (smooth curve) corresponds to the NBD.

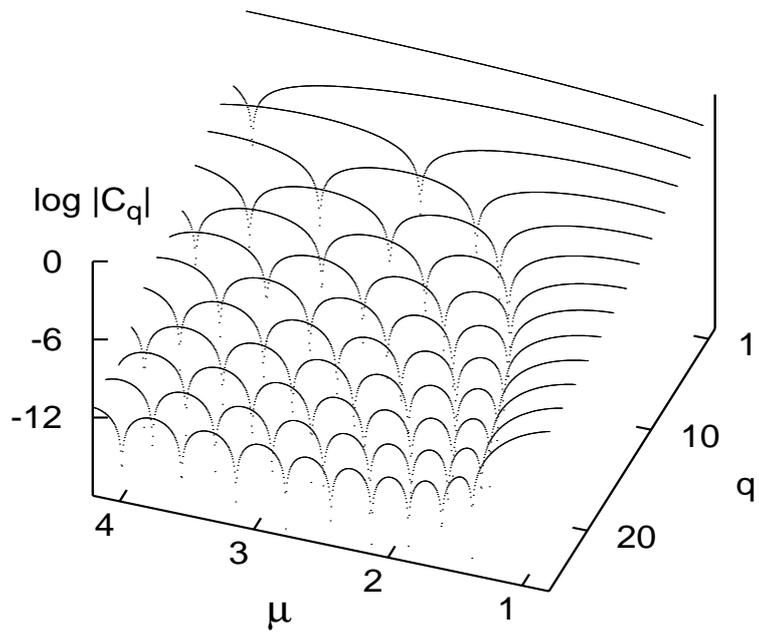
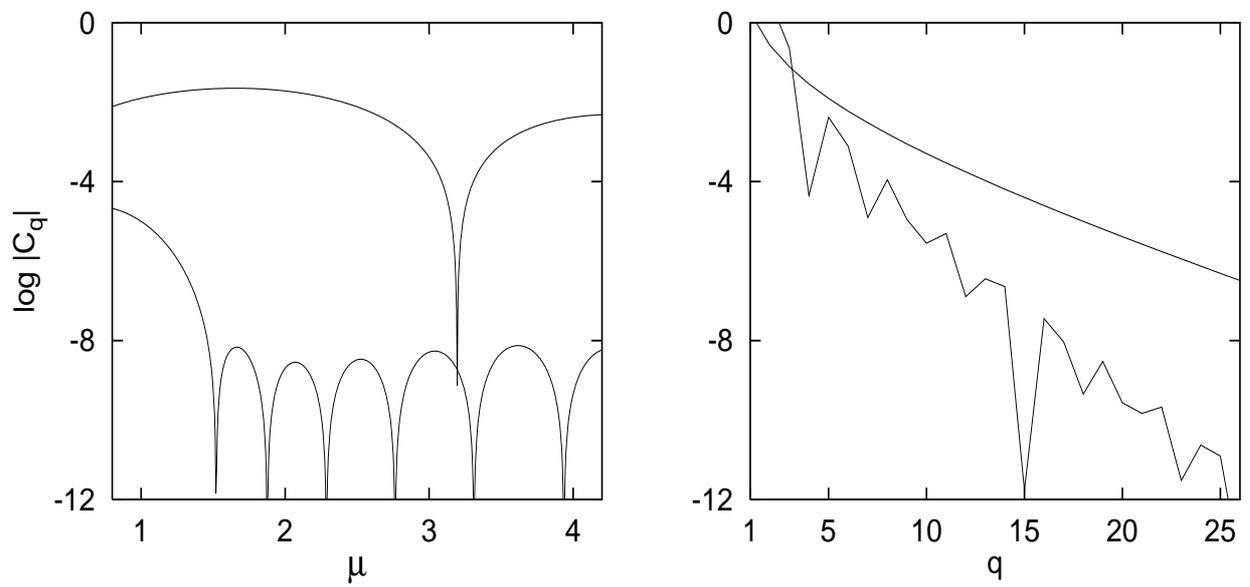
\end{document}